\documentclass{jpp}

\usepackage{graphicx}
\usepackage{natbib}

\begin{document}

\title[Acceleration by parallel Alfv\'en waves]{Plasma acceleration by the interaction of parallel propagating Alfv\'en waves }

\author[F.~Mottez]{F.~MOTTEZ \thanks{Email address for correspondence: fabrice.mottez@obspm.fr}}

\affiliation{Laboratoire Univers et Th\'eories (LUTH), Observatoire de Paris; CNRS UMR8102, Universit\'e Paris Diderot; 5 Place Jules Janssen, 92190 Meudon, France}







\maketitle

\begin{abstract}
It is shown that two circularly polarised Alfv\'en waves that propagate along the ambient magnetic field in an uniform plasma trigger non oscillating electromagnetic field components when they cross each other.  The non-oscilliating field components can accelerate ions and electrons with great efficiency. This work is based on particle-in-cell (PIC) numerical simulations and on analytical non-linear computations. The analytical computations are done for two counter-propagating monochromatic waves. The simulations are done with monochromatic waves and with wave packets. The simulations show parallel electromagnetic fields consistent with the theory, and they show that the particle acceleration result in  plasma cavities and, if the waves amplitudes are high enough, in ion beams. These acceleration processes could be relevant in space plasmas. For instance, they could be at work in the auroral zone and in the radiation belts of the Earth magnetosphere. In particular, they may explain the origin of the deep plasma cavities observed in the Earth auroral zone.
\end{abstract}


\section{Introduction}


We investigate plasma acceleration based on the non-linear interaction of counter-propagating MHD waves. 
A first analysis was presented in \citet{M2012}. The present paper is devoted to a more thorough analysis of this process, that is now called Acceleration by Parallel Alfv\'en Waves Interactions (APAWI).

 Most of the known Alfv\'enic acceleration mechanisms  involve short perpendicular scales \citep{Knudsen_2001,mottez_2012b}. For instance $k_\perp c/\omega_{pe} \sim 1$ \citep{Goertz_1984,Thompson_1996,Lysak_2003} or $k_\perp \rho_i \sim 1$ \citep{Hasegawa_1978} where $\rho_i$ is the ion Larmor radius and $k_\perp$ is the wavelength in the direction orthogonal to the ambient magnetic field.
 At odds with all these processes, APAWI involves waves in parallel propagation to the magnetic field $(k_\perp=0)$.
 Because the magnetic field direction is a privileged direction of propagation of Alfv\'en waves,  APAWI is potentially relevant in space plasma physics. The first paper was developed in connexion with acceleration in the Earth auroral zone. But it could be at work in any place where trapped or reflected Alfv\'en waves counter-propagate; for instance in the Earth radiation belts, or in closed magnetic loops in the solar corona. 
 

 The investigation of APAWI in \citep{M2012} was initially based on numerical simulations and completed with theoretical computations.
In  \citep{M2012}, the numerical simulations put in evidence the settlement of non oscillatory parallel electric field (if the waves polarisation are appropriate), its bilinear dependence on the waves amplitudes, and the creation of plasma cavities. This last property was interpreted as a proxy of plasma acceleration.
A theoretical computation of the electric field parallel to $\vec B_0$ confirmed the existence of the non oscillatory component seen in the numerical simulations. 
 
Both the numerical and the theoretical aspects of the problem are re-analysed here.
The following points are studied more thoroughly. 

A series of simulations has been conducted to complement those presented in \citet{M2012}. 
The simulations in the previous paper were conducted with a PIC code that computes electron guiding centre and  full ion dynamics \citep{Mottez_1998}. Did this code contain approximations that could have biased the physics ? In order to answer that question, the simulations presented in the present paper are done with another PIC code where the electron full dynamics is taken into account. This code also allows simulations with lower ambient magnetic fields. 
The simulation are run over longer durations. 
All these differences are used to test the robustness of the acceleration mechanism presented in \citep{M2012}. 

{The theory developed in  \citep{M2012} showed the same wave interaction terms as in the simulations; but it showed also a few supplementary terms that did not appear in the simulations. These terms were proportional to the square of the amplitude of a \textit{single} wave. Therefore, they were not associated with wave-wave interaction. We called them self-wave coupling terms and the drawback that they represented was not resolved in \citep{M2012}.} In section \ref{sec_theorie} and appendix \ref{sec_faraday}, we derive again the computation of the electric field along $\vec B_0$ developed in  \citep{M2012}, with the addition of the constraint associated with the Maxwell-Faraday law. It then appears that the self-wave-coupling terms disappear, and we obtain exactly what is seen in the numerical simulations. 

Because of this reconsideration of the theoretical work, the reader can use the present paper as the fundamental paper describing the principle of APAWI. All the valid theory developed up to now is here. The former paper \citep{M2012} can be considered as a seminal paper. It is still interesting because of the parametric study conducted with numerical simulations (it is not repeated here) that allows for both qualitative and quantitative tests of the theory.  

It might seem impossible that an electric field that is locally perpendicular to the magnetic field $\vec B$ can accelerate particles in its average direction $\vec B_0$. Nevertheless, acceleration occurs. 
The ability of APAWI to accelerate electrons and ions is demonstrated with simple analytical calculations in section \ref{sec_acceleration}. 

A series of signatures that could help to characterise APAWI in observational data is presented in section \ref{sec_signature}. 

In section \ref{sec_examples_space}, a few examples of published observations where APAWI might be relevant are discussed.

The main conclusions of our study as well as a short discussion relative to other theories involving parallel Alfv\'en waves interaction are presented in the last section.

\section{Numerical method and simulation parameters} \label{simulation_parameters}


The physical variables are reduced to dimensionless variables. Time and frequencies are
normalized by the electron plasma frequency $\omega_{p0}$ that correspond to a reference background 
electron density $n_0$. Velocities are normalized to the speed of light $c$, and the magnetic
field is given in terms of the dimensionless electron gyrofrequency $\omega_{ce}/\omega_{p0}$.
The mass unit is the electron mass $m_e$. Therefore, the units   (starting from the Maxwell Eq. in the MKSA system) are 
$c/\omega_{p0}$ for distances, $\omega_{p0}/c$ for wave vectors,  $e$ for charges,
$e n_0$ for the charge density, $c\omega_{ce}/\omega_{p0}$  for 
the electric field.
In the following parts of this paper,  numerical values and  figures are expressed in this system of units.

For the understanding of APAWI, it is necessary to use simultaneously two systems of coordinates. 
The coordinates system $(X,Y,Z)$ is associated with the orthonormal basis $(\vec{e_X},\vec{e_Y},\vec{e_Z})$ where $\vec{e_X}$ is parallel to $\vec{B_0}$.
The unit vectors $\vec{e_Y},\vec{e_Z}$ are set arbitrarily, as long as they contribute with $\vec{e_X}$ to form an orthonormal basis. 
In the approximation of a uniform magnetic field $\vec{B_0}$ applied in the computation, this coordinate system is time invariant and uniform.
We can attach a local frame of reference to the local magnetic field $\vec{B}(X,t)$. The $\vec{e_x}$ axis of this local frame of reference of coordinates $(x,y,z)$ is parallel to the local magnetic field. For convenience, the basis $(\vec{e_x},\vec{e_y},\vec{e_z})$ can be defined as the result of the rotation $R_Z$ of axis $\vec{e_Z}$ and angle $\theta_Y$ of the set $(\vec{e_X},\vec{e_Y},\vec{e_Z})$,  followed by a rotation $R_{Y1}$ of axis $\vec{e_{Y1}}= R_Z(\vec{e_y})$ and angle $-\theta_Z$.

This paper is based on a series of numerical simulations, whose {naming convention} is AWC (for Alfv\'en wave crossing) followed by a number, or a number and a letter. These simulation share most of their characteristics with AWC009, that is therefore used as a reference simulation. 

 In AWC009, two wave packets propagate in a uniform plasma. 
 There are    1638400 particles of each species in the simulation;
 it corresponds to         100 particles per cell.
 There are  65536 time steps, defined by $\Delta t = 0.1$,
 corresponding to a time lapse $ t_{max}= 6553.6$.
 The size of the whole simulation domain is $ 4096\Delta x \times  4\Delta y$.
 where $\Delta x=\Delta y=  0.10$ is the size of the grid cells.
  The electron thermal velocity is 
 $v_{te}= 0.1$; the ion and electron temperatures are the same. 
 The ion to electron mass ratio is reduced to $m_i/m_e= 400.$.
Each wave packet is initialized as the sum of  8 sinusoidal waves with their maximum 
at $X=  300$ for the first wave packet, and $X= 100$ for the second wave packet. 
The equations defining the waves polarisations are given in the appendix in \citep{mottez_2008_a}.
  The first wave packet propagates downward (in the direction of decreasing values of $X$). 
  The sinusoidal waves that compose the wave packet have the same amplitude, given by the wave magnetic field 
 $\delta B  = 0.05\times B_0$.
  The waves have a right handed circular polarization.
 Their wavelengths are $\lambda= \lambda_0/m$ where $\lambda_0=409.6$
 and $m$ varies from 1 to  8. 
 The phase velocities vary from  0.109 for the shortest wave  
 to  0.046 for the longest. 
 (The Alfven velocity is $V_A=  0.04$.) 
  The second wave packet propagates upward. Apart from its direction of propagation and location, 
  its characteristics are the same. 

\begin{figure}
\vspace*{2mm}
\begin{center}
\includegraphics[width=8.3cm]{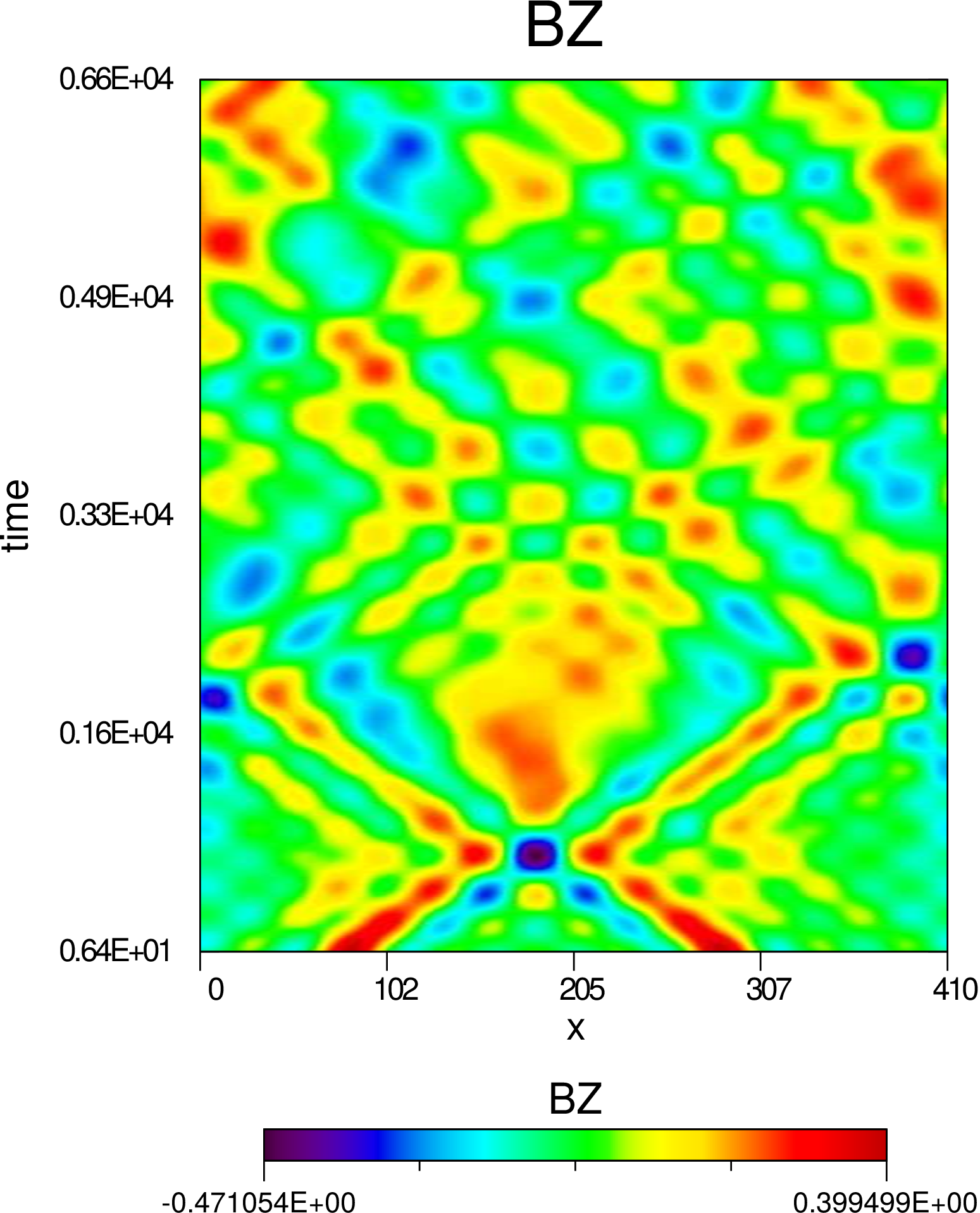} 
\end{center}
\caption{AWC009. Temporal stack plot of the transverse magnetic field $B_Z(X,t)$.}
\label{fig_AWC009_CBZ}
\end{figure}

  {The $B_z$ component of the magnetic field is a proxy of the wave packets propagation. It is plotted as a function of $X$ and time on Fig. \ref{fig_AWC009_CBZ}.
   We can see that the two packets cross each other at time $t \sim 10^4$ and $X \sim 200$, and latter a times $t \sim 2. 10^4$, $t \sim 3. 10^4$ etc. }

   For the other simulations, a few parameters are varied. When a physical parameter is varied (for instance, a wave amplitude), the number is different. 
{The parameters of these simulations are summarized in Table \ref{table_simulations}. We can notice that these simulations all involve the right-hand circularly polarized waves propagating in opposite directions, as in AWC009. We made the deliberate choice to concentrate on this subset of parameters because, as shown in \citep{M2012}, it represent a favourable case for parallel to $B_0$ electric field, and because the RH waves are less dispersive than the LH ones. This make the physical interpretation of the simulation easier. Concerning dimensions, a 2D simulation was presented in \citep{M2012} 
with the same initial configuration as the 1D simulations. The plasma behaved exactly as in 1D simulations. 
This is why 2D simulations are not repeated in the present study.} 

The simulation AWC009 is based on a magnetic field $B_0= 0.8$ that is five times smaller than those in \citep{M2012}. This is a deliberate choice in order to check the robustness of APAWI relative to the ambient magnetic field amplitude. 
{Nevertheless, we can notice that the plasma $\beta$ (see Table \ref{table_simulations}) is typical of a highly magnetized plasma ($\beta << 1$). In the simulations with $B_0=4$,  $\beta < m_e/m_i=10^{-2}$ is typical of the largest part of acceleration region of the auroral zone. To illustrate more precisely the physical context, let us notice that the "classical" theories of acceleration by oblique Alfv\'en waves (therefore not APAWI)  involve inertial AW in this range of altitudes \citep{Thompson_1996,Lysak_2003}.  With $B_0=2$ or $0.8$, $m_e/m_i < \beta <<1$ is representative of the upper part of the auroral zone where acceleration by oblique AW is caused by  kinetic AW \citep{Hasegawa_1978}. 
Such low values of the plasma $\beta$ can also be relevant for the study of the inner solar corona.
} 

{Considering $\beta <<1$ reinforces the choice of simulations with right-hand (RH) circularly polarized waves rather than left-hand (LH) waves, because in this regime, as shown by \cite{Buti_2000}, LH Alfv\'en wave packets are unstable to collapse or with possible change of polarization, while RH Alfv\'en wave packets are stable. Such phenomenon would make the interpretation of the simulations more difficult relatively to APAWI.}

  We can notice that because of the limited size of the simulation domain, and hence of the wavelengths, the wave packet is dispersive. The frequencies corresponding to the small wavelengths is higher than 
  the ion gyro-period ($T_{ci}\omega_{pe}=3141$), placing those waves nearly in the domain of 
  electromagnetic whistler waves rather than MHD Alfv\'en waves. Nevertheless, these waves are on the same branch of the linear dispersion relation, and they are both observed in the auroral zone and many other regions of the magnetosphere.

   \begin{table*}
   \begin{tabular}{|l|c|c|c|c|c|c|l|}
   \hline
   name  & $B_0$ & $\Delta t$& $\delta B/B_0$  &  $\beta$ & $\lambda$ & \\   \hline
 AWC001 &  0.8 &  0.2  & 0.02   & $1.6 \, 10^{-2}$ &      204.8     & unperturbed $f_e$ and $f_i$ \\
 AWC002 &  0.8 &  0.2  & 0.2  &  $1.6 \, 10^{-2}$&      204.8     & weak modulation of $f_e$ and $f_i$\\
 AWC003 &  0.8 & 0.2   & 0.02  &  $1.6 \, 10^{-2}$ &  51.2-409.  & weak modulation of $f_e$ and $f_i$\\
 AWC004 &  0.8 & 0.1   & 0.04   &  $1.6 \, 10^{-2}$&      204.8    & modulation of  $f_e$ and $f_i$\\
 AWC005 &  0.8 & 0.1   & 0.024 &  $1.6 \, 10^{-2}$ &  51.2-409. & modulation of $f_e$ and $f_i$\\
 AWC006 &  2.0 & 0.05  & 0.10   &  $0.5\, 10^{-2}$ &      204.8     & modulation of $f_e$ and $f_i$\\
 AWC007 &  2.0 & 0.05  & 0.06   & $0.5\, 10^{-2}$ & 51.2-409.  & $f_e$ vortices,  $f_i$ vortices \& beams\\
 AWC008 &  0.8 & 0.1   &  0.10  & $1.6 \, 10^{-2}$ &      204.8    & electron cavities, ion beams\\
 AWC009 &  0.8 & 0.1   &  0.05  & $1.6 \, 10^{-2}$ &  51.2-409.  & electron cavities \& bumps, ion beams.\\
 AWC014 &  4.0 & 0.025 & 0.20  & $0.25 \, 10^{-2}$ &      204.8     & modulation of $f_e$, ion beams\\
 AWC015 &  4.0 & 0.025 & 0.12   & $0.25 \, 10^{-2}$ &  51.2-409.   & electron vortices \& cavities\\
        &      &       &            &       &                                   &  bulk ion acceleration \& beam\\
 AWC016 &  4.0 & 0.025 & 0.05   & $0.25 \, 10^{-2}$ &      204.8     & $f_e$ and $f_i$ hardly not modulated\\
 AWC017 &  4.0 & 0.025 & 0.02   & $0.25 \, 10^{-2}$ &  51.2-409.   & $f_e$ and $f_i$ hardly not modulated\\
 AWC019 &  4.0 & 0.025 & 0.06   & $0.25 \, 10^{-2}$ &  51.2-409.   &  strong electron cavities, ion beams \\
   \hline
   \end{tabular}
   \caption{Varied simulation parameters and comments on the evolution of the particle distribution functions $f_e(X,V_x)$ (electrons) and $f_i(X,V_x)$ (ions). The wave amplitudes $\delta B/B_0$ and the range of wavelengths $\lambda$ are the same for the two waves/wave packets; they are not repeated in the table.}
   \label{table_simulations}
   \end{table*}


%


\begin{figure}
\vspace*{2mm}
\begin{center}
\includegraphics[width=8.3cm]{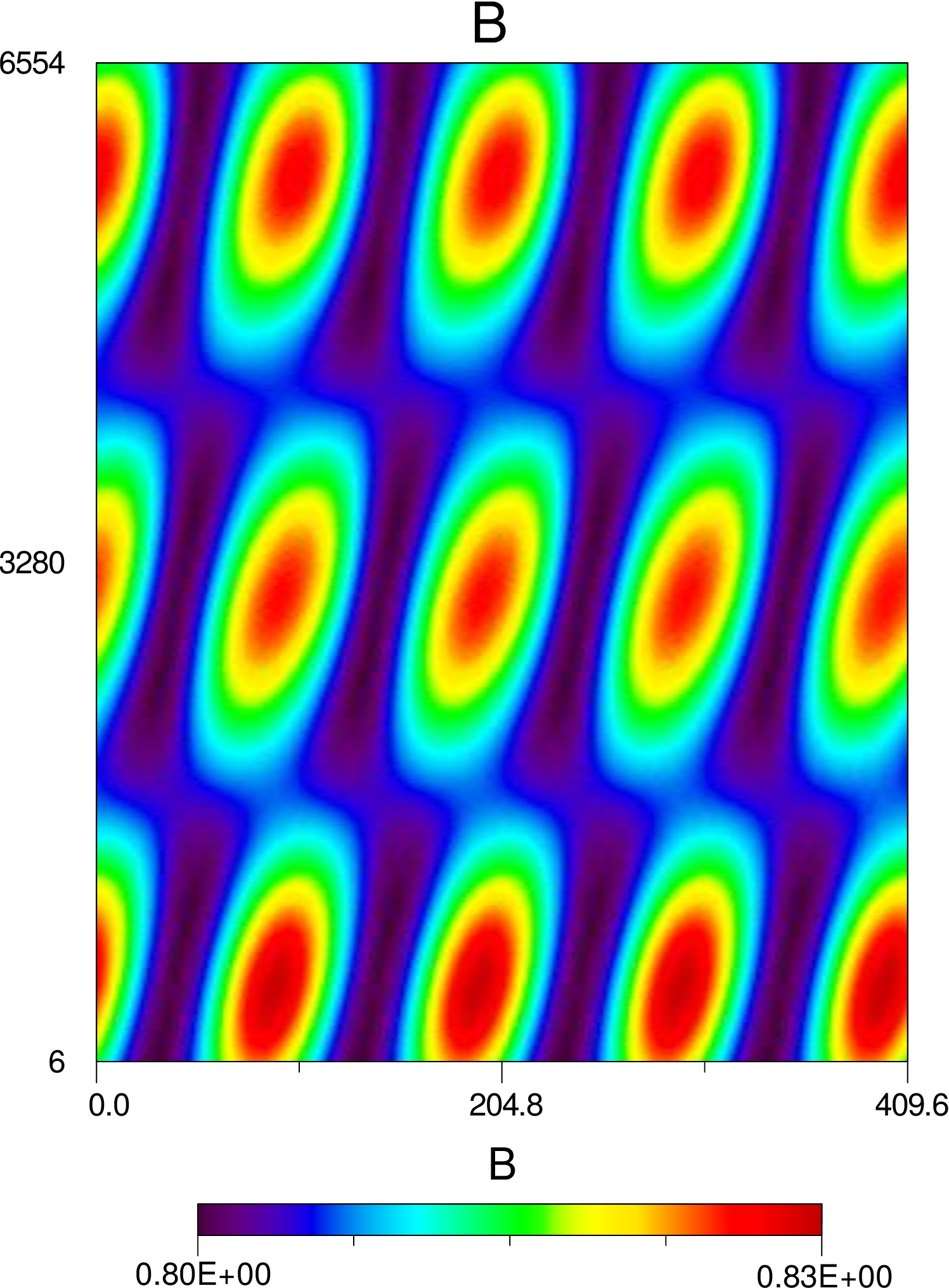} 
\end{center}
\caption{AWC008. Temporal stack plot of the magnetic field amplitude $B(X,t)$.}
\label{AWC008_CBMOD}
\end{figure}

\begin{figure}
\vspace*{2mm}
\begin{center}
\includegraphics[width=8.3cm]{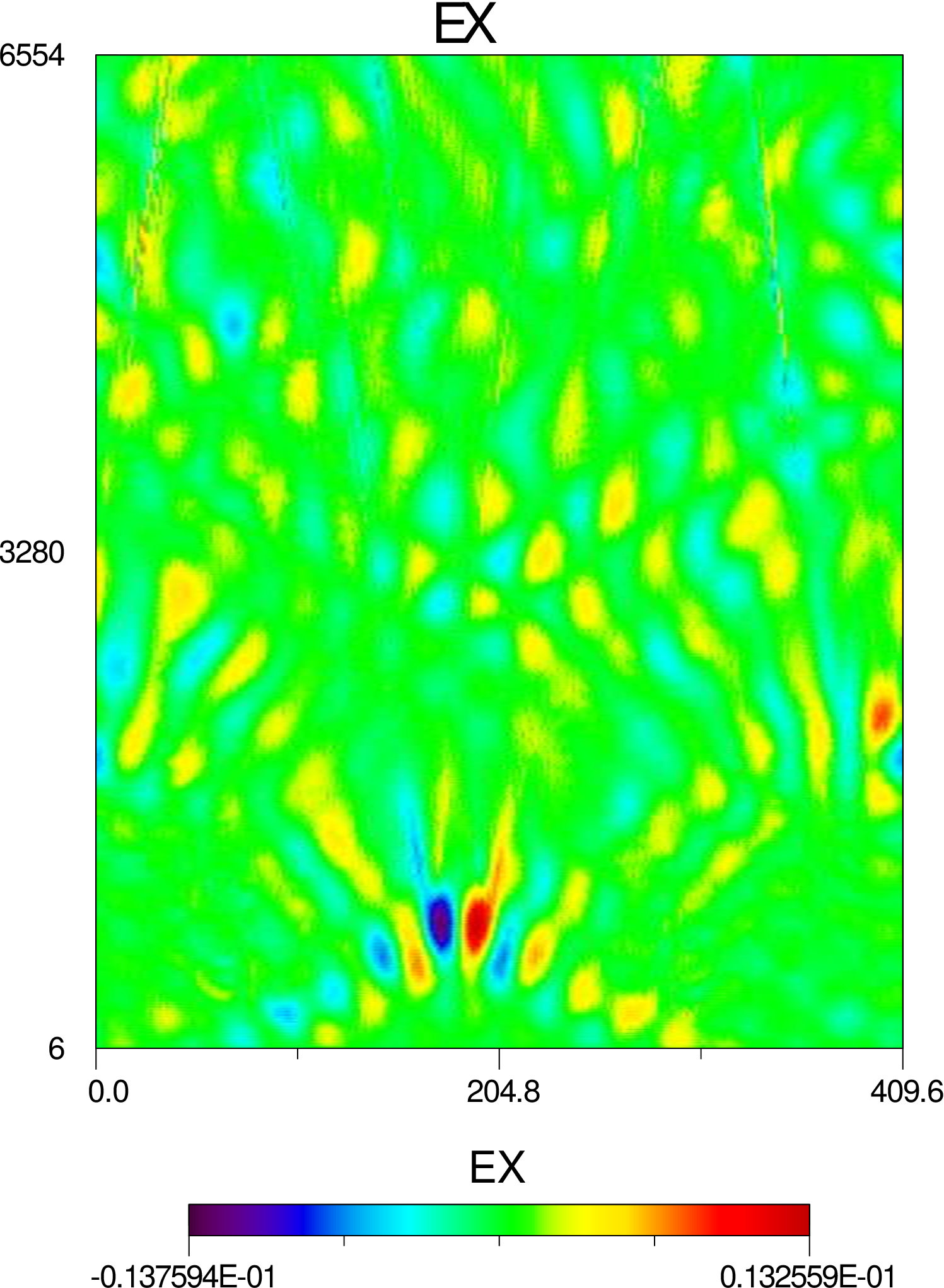} 
\end{center}
\caption{AWC009. Temporal stack plot of the electric field  $E_X(X,t)$ parallel to $\vec{B}_0$. In order to remove plasma oscillations near the plasma frequency, the signal was filtered with a first order low-pass filter (cut-off frequency $0.5 \omega_{pe}$).}
\label{fig_AWC009_CEX}
\end{figure}

\begin{figure}
\vspace*{2mm}
\begin{center}
\includegraphics[width=8.3cm]{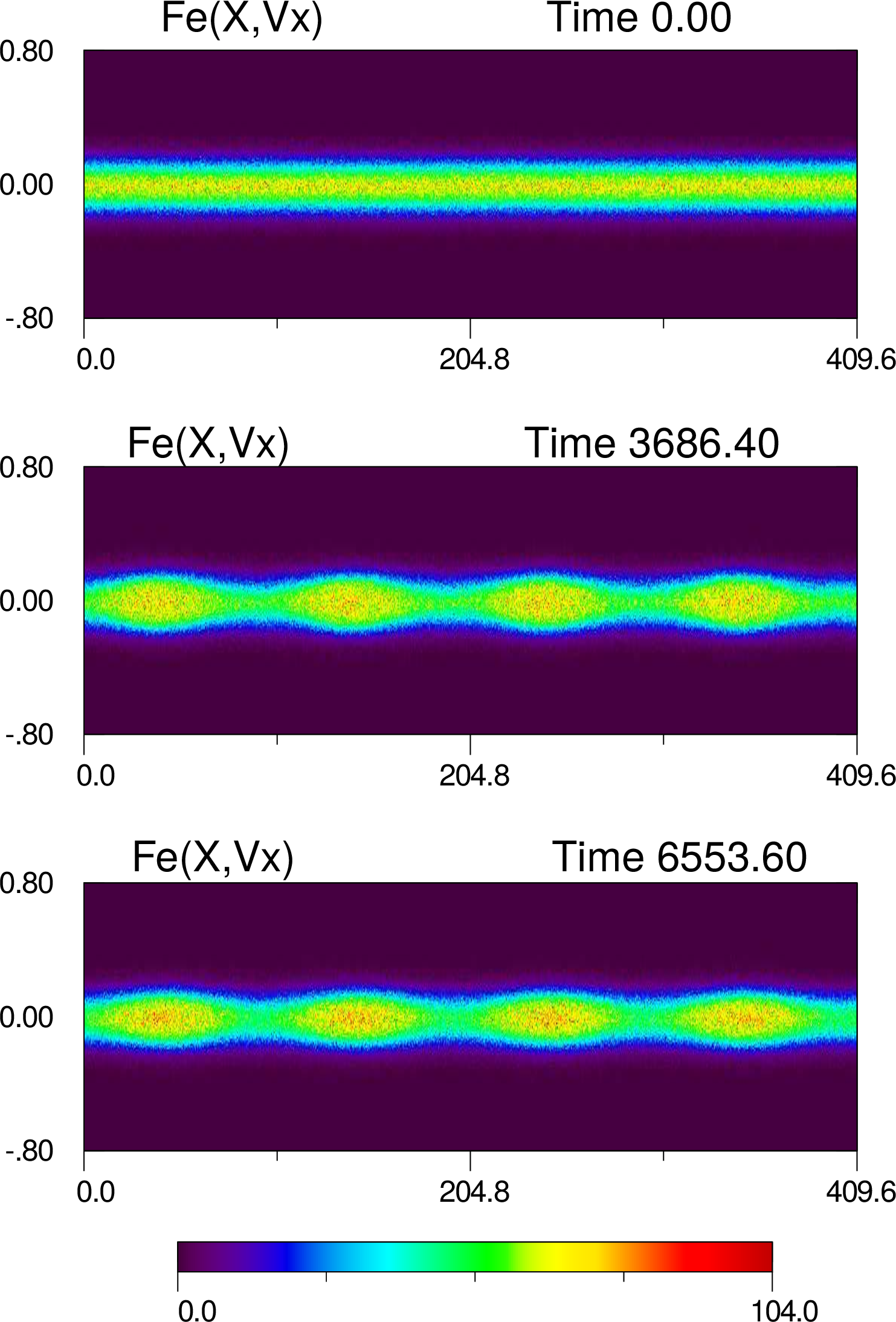} 
\end{center}
\caption{AWC008. Snapshots of the electron distribution function as a function of $X$ and the velocity $V_x$ parallel to the local magnetic field $\vec B$.}
\label{fig_AWC008_FeXVPAR}
\end{figure}

\begin{figure}
\vspace*{2mm}
\begin{center}
\includegraphics[width=8.3cm]{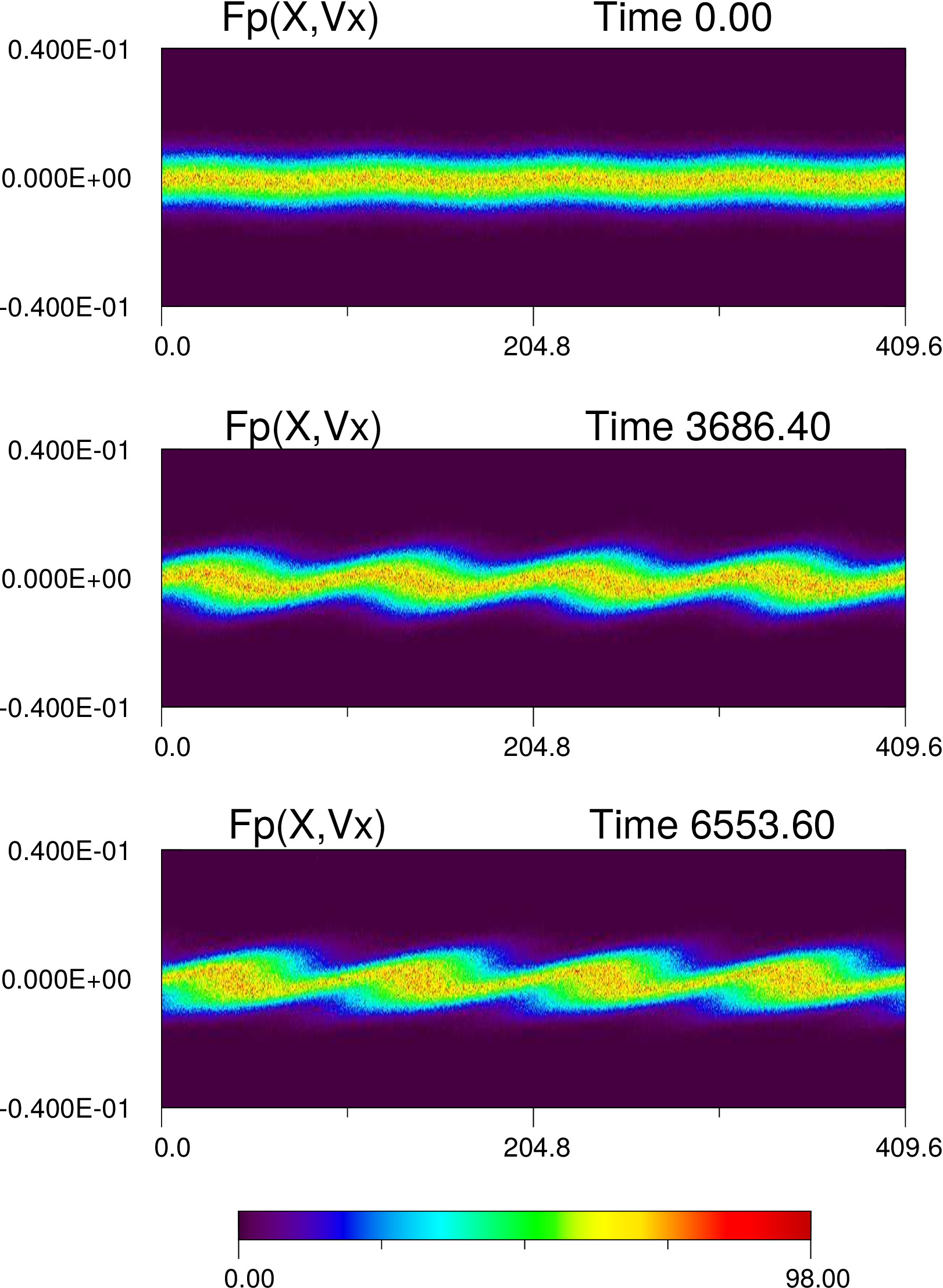} 
\end{center}
\caption{AWC008. Snapshots of the ion distribution function as a function of $X$ and the velocity  $V_x$ parallel to the local magnetic field $\vec B$.}
\label{fig_AWC008_FpXVPAR}
\end{figure}

\begin{figure}
\vspace*{2mm}
\begin{center}
\includegraphics[width=8.3cm]{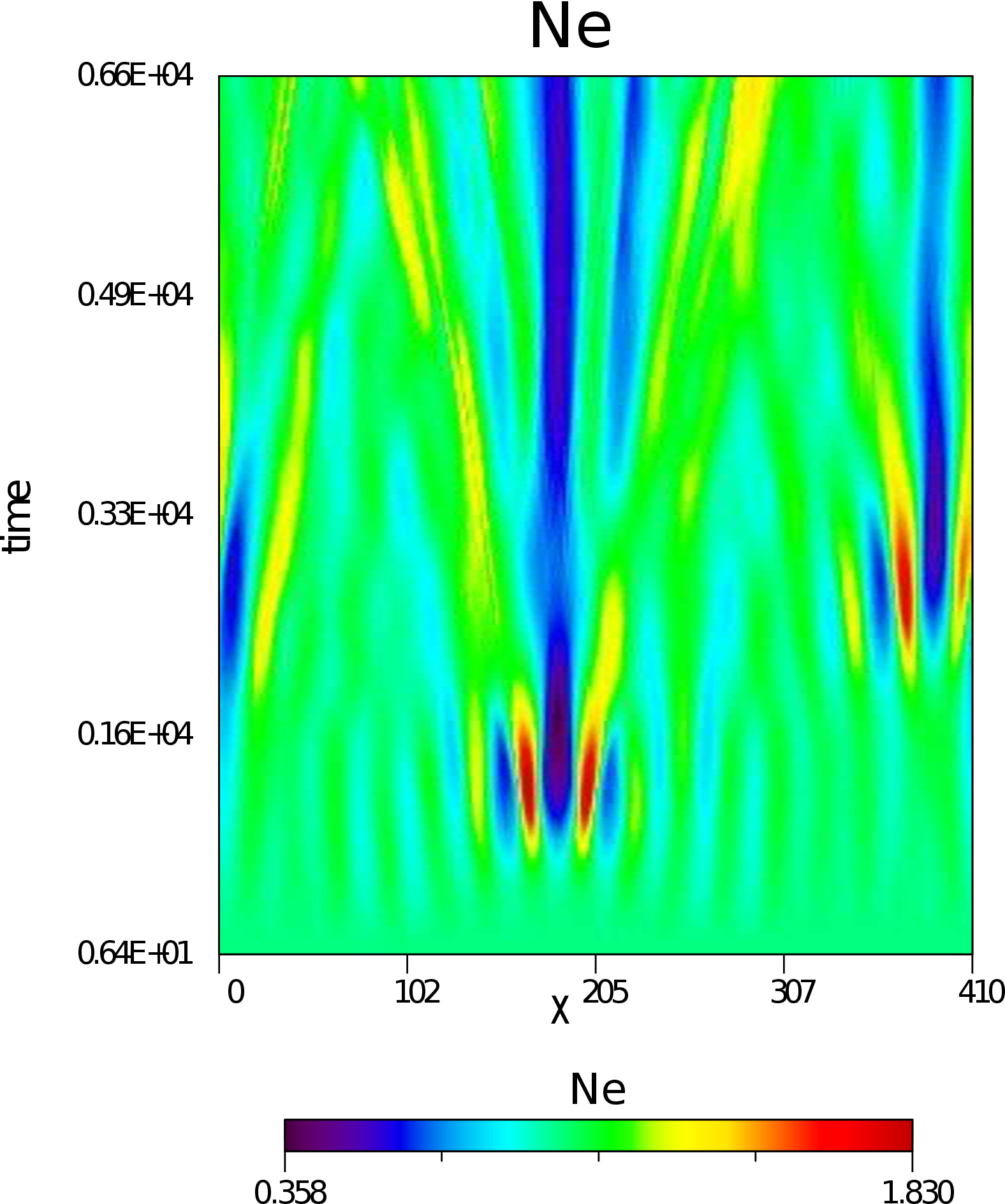} 
\end{center}
\caption{AWC009. Temporal stack plot of the electron density  $N_e(X,t)$.}
\label{fig_AWC009_CNE}
\end{figure}

\begin{figure}
\vspace*{2mm}
\begin{center}
\includegraphics[width=8.3cm]{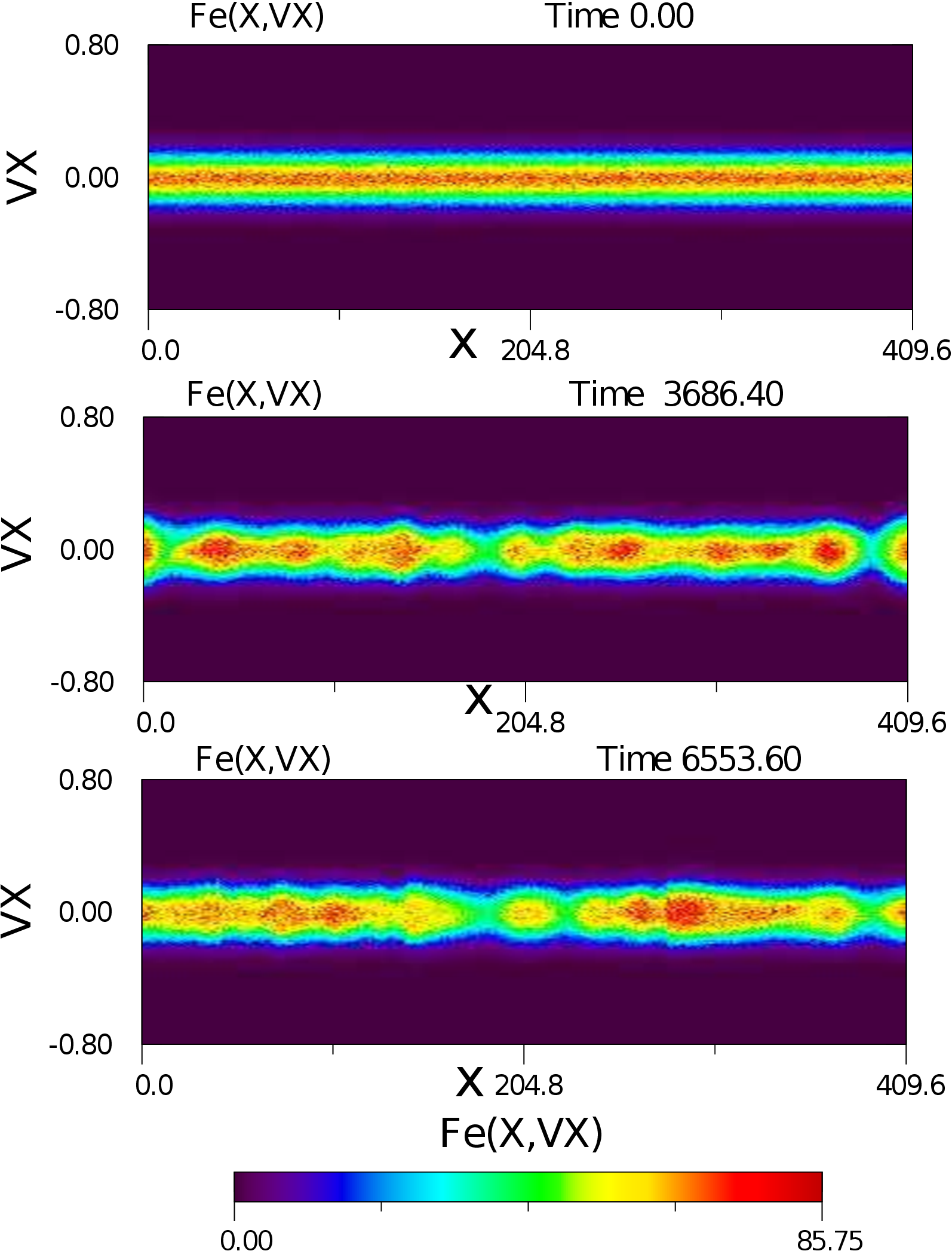} 
\end{center}
\caption{AWC009. Snapshots of the electron distribution function as a function of $X$ and of the velocity $V_X$ parallel to the \textit{mean magnetic field} $\vec{B}_0$.}
\label{fig_AWC009_FeXVX}
\end{figure}

\begin{figure}
\vspace*{2mm}
\begin{center}
\includegraphics[width=8.3cm]{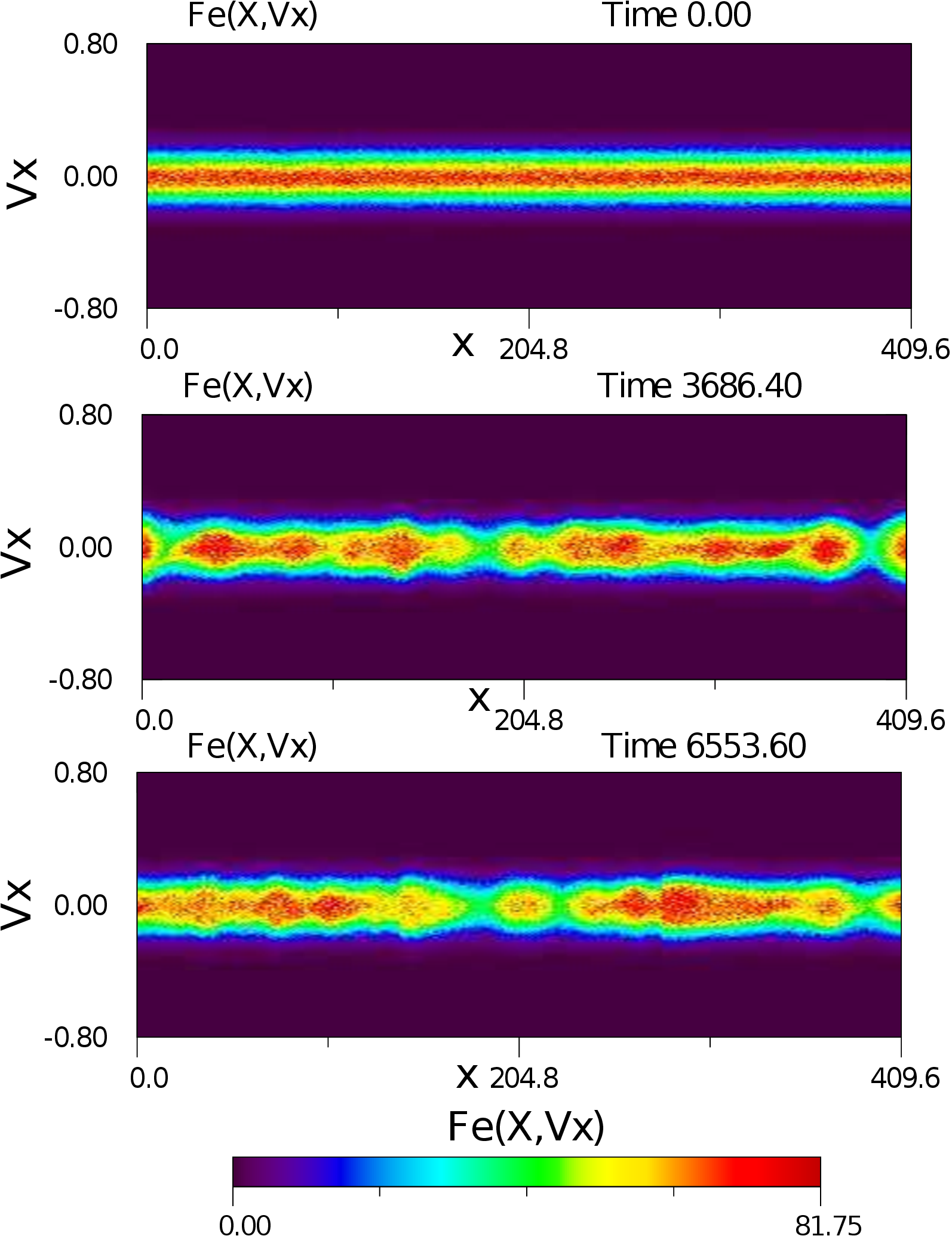} 
\end{center}
\caption{AWC009. Snapshots of the electron distribution function as a function of $X$ and the velocity $V_x$ parallel to the \textit{local magnetic field} $\vec B$.}
\label{fig_AWC009_FeXVPAR}
\end{figure}

\begin{figure}
\vspace*{2mm}
\begin{center}
\includegraphics[width=8.3cm]{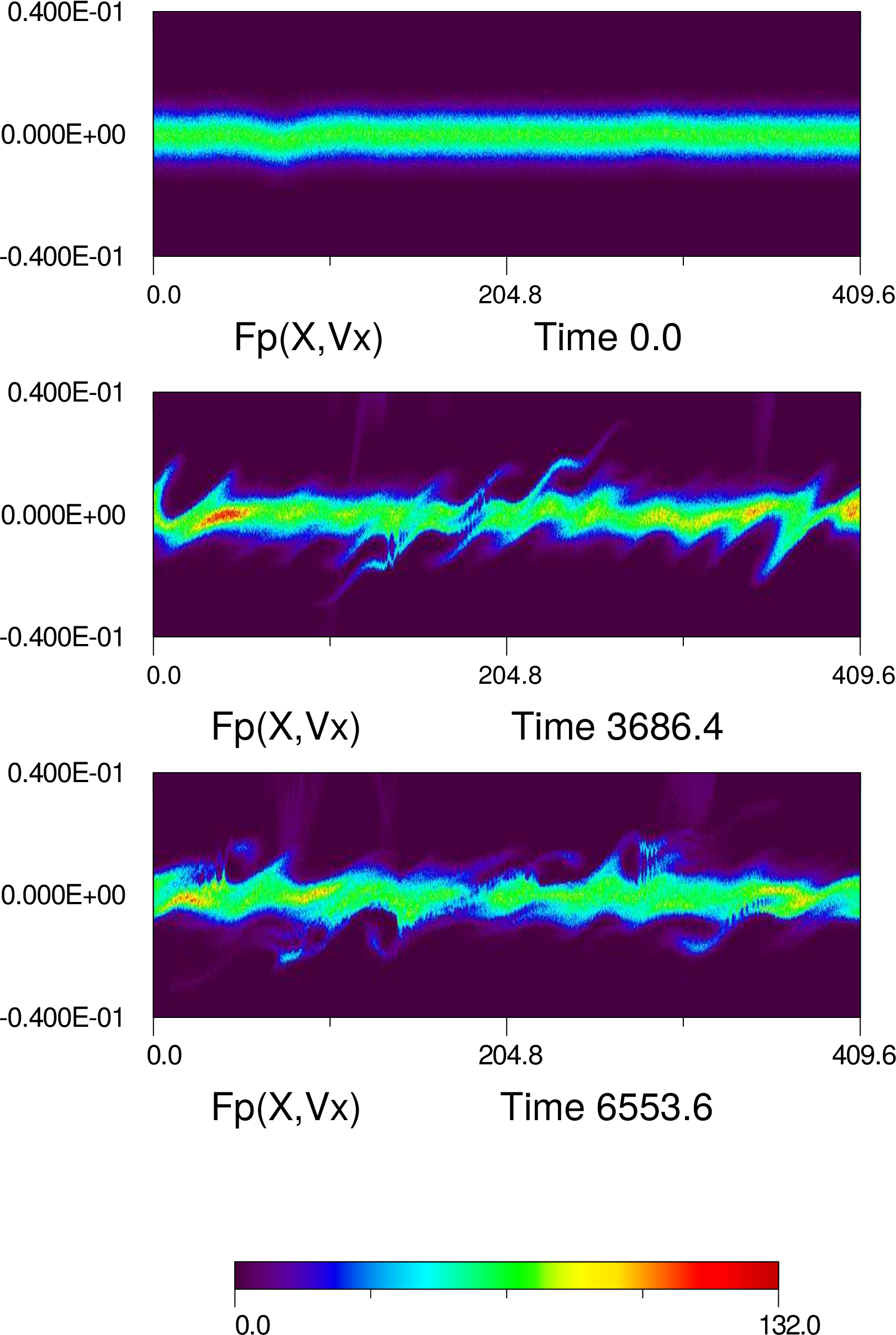} 
\end{center}
\caption{AWC009. Snapshots of the ion distribution function as a function of $X$ and the velocity $V_x$ parallel to the \textit{local magnetic field} $\vec B$.}
\label{fig_AWC009_FpXVPAR}
\end{figure}

\section{Properties of the electromagnetic field} \label{sec_theorie}

We consider two sinusoidal waves labelled 1 and 2, and their associated magnetic field,
\begin{eqnarray} 
B_X &=& B_0 \label{eq_B}\\
\nonumber
B_Y &=& B_{1Y} \cos(\omega t - kX +\phi_{B_{1Y}})+
B_{2Y} \cos(\omega t + kX +\phi_{B_{2Y}})\\   
\nonumber
B_Z &=& B_{1Z} \cos(\omega t - kX +\phi_{B_{1Z}})+
B_{2Z} \cos(\omega t + kX +\phi_{B_{2Z}}) 
\end{eqnarray}
We write the phase relations in the form 
\begin{equation}
\phi_{B_{1Z}} = \phi_{B_{1Y}} + {n_1 \pi}/{2}
\end{equation}
with $n_1 =\pm 2$ for linear polarization, $n_1=+1$ for right-handed waves, 
$n_1=-1$ for left-handed waves. With circularly polarised waves, $B_{1Y}=B_{1Z}$, $B_{2Y}=B_{2Z}$ and we write simply $B_1$ and $B_2$.
Because the magnetic field modulus $B$ is required in the next section to compute the dynamics of the particles, 
we develop it here,
\begin{eqnarray} \label{module_B}
B^2= B_0^2
&&+B_{1Y}^2 \cos^2(\omega t - kX +\phi_{B_{1Y}})
\\ \nonumber
&&+B_{2Y}^2 \cos^2(\omega t + kX +\phi_{B_{2Y}})
\\ \nonumber
&&+B_{1Z}^2 \cos^2(\omega t - kX +\phi_{B_{1Y}}+n_1 \frac{\pi}{2})
\\ \nonumber
&&+B_{2Z}^2 \cos^2(\omega t + kX +\phi_{B_{2Y}}+n_2 \frac{\pi}{2})
\\ \nonumber
&&+ B_{1Y} B_{2Y} \left[ \cos(2 \omega t + \phi_+) + \cos(2kX +\phi_-) \right] 
\\ \nonumber
&&+ B_{1Z} B_{2Z}  \cos(2 \omega t + \phi_+ + (n_1+n_2)\frac{\pi}{2}) 
\\ \nonumber
&&+ B_{1Z} B_{2Z}   \cos(2kX +\phi_- + (n_2-n_1)\frac{\pi}{2}),
\end{eqnarray} 
where $\phi_{-}=\phi_{B_{2Y}}-\phi_{B_{1Y}}$ and $\phi_{+}=\phi_{B_{2Y}}+\phi_{B_{1Y}}$.
We find the well known property that a single wave with circular polarisation has a uniform 
magnetic field amplitude. 
With two waves with the same circular polarisation ($n_1=n_2$), the modulus is purely space dependent, 
\begin{equation} \label{module_B_same_circular}
B^2= B_0^2 +B_1^2+B_2^2 +2B_1 B_2 \cos(2kX +\phi_-).
\end{equation}
Fig \ref{AWC008_CBMOD} 
shows the stack plot of the magnetic field modulus in simulation AWC008. We can see that unlike Eq. (\ref{module_B_same_circular}), it is not purely space dependent. This may be caused by an imperfect initialisation of the waves polarisation. Nevertheless, it is not oscilliating either and the regions of maximas and of minimas of $B$ are static. 
Two waves with opposite circular polarisation ($n_1=-n_2$) have a purely time dependent modulus, 
\begin{equation} \label{module_B_opposite_circular}
B^2= B_0^2 +B_1^2+B_2^2 +2B_1 B_2 \cos(2 \omega t +\phi_+).
\end{equation}

The ability of the two waves to create an electric field with a component $E_X$ that is parallel to the average magnetic field was investigated in \citep{M2012}.
 This property holds with the simulations made since then. But it was found in the theoretical derivation in \citep{M2012} that some terms appeared that could not be seen in the numerical simulations.They were self-coupling terms  in the sense that they involved only one wave. A new derivation of $E_X$ is presented in appendix \ref{sec_faraday}, where the constraint brought by the Ampere-Faraday law is taken into account. We find a parallel electric field $E_X$ that contains only the terms seen in the 
numerical simulations, and not the self-coupling ones,
\begin{eqnarray} \label{EX_general}
&& {k B_0 E_X}/{\omega} =\\ \nonumber  
 \nonumber
&&  - B_{1Z} B_{2Y} [\cos{(2kX +\phi_{-}-\frac{n_1  \pi}{2})}
  +\cos{(2 \omega t +\phi_{+}+\frac{n_1 \pi}{2})}]    
  \\  \nonumber & &
  + B_{1Y} B_{2Z}[\cos{(2kX +\phi_{-}+\frac{n_2  \pi}{2})}
  +\cos{(2 \omega t +\phi_{+}+\frac{n_2 \pi}{2})}].
\end{eqnarray}

When the two waves are linearly polarized and coplanar, $E_X=0$. 
When the two waves are linearly polarized and orthogonal, for instance
$\phi_{B_{1Z}}=\phi_{B_{2Y}}=0$, 
$E_X= 2  B_{1Y} B_{2Z} \cos{(\omega t - kX +\phi_{B_{1Y}})} \cos{(\omega t + kX +\phi_{B_{2Z}})}$.

When the two waves are circularly polarized of opposite direction (left and right-handed),
$E_X$ is purely temporal of period $2 \omega$, there is no spatial dependence. 

When the two waves are circularly polarized, with the same polarization ($n_1=n_2$)
$E_X$ is purely spatial of wave vector $2k$, there is no temporal dependence.
\begin{equation} \label{EX_general_same_circular}
 {k B_0 E_X}/{\omega} =2 B_1 B_2  \cos(2kX+\phi_- + n \frac{\pi}{2}).
\end{equation}
Eq. (\ref{EX_general}) fits the electric field $E_X$ found  with the simulations presented in \citep{M2012} as well as the simulations attached to the present work. This point was analysed in details in \citep{M2012} and we do not repeat the analysis here.

It is interesting to see that $B$ and $E_X$ are purely space dependent under the same circumstances (two waves with the same circular polarisation) and purely time dependent as well  (two waves with opposite circular polarisations).

{This result, set analytically for two sinusoidal waves holds for two wave packets. The stack plot of $E_X$ is shown in Fig. \ref{fig_AWC009_CEX} 
 for simulation AWC009. We can see the setting of a finite $E_X$ at the places and times that correspond to wave packets crossings. This electric field is stationary... as long as the crossing lasts. Later, $E_X$ vanishes.}

\section{Particle acceleration along the mean magnetic field direction} \label{sec_acceleration}

Let us consider a particle of velocity $\vec V = \vec V_x+\vec V_\perp$ where
the indexes $\parallel$ and $\perp$ are relative to the local magnetic field direction.
We note $\vec b = \vec B / B$. Considering the guiding centre theory of motion, the motion along the magnetic field line is governed by the parallel electric field $E_x$ and 
the parallel gradient of the magnetic field modulus 
\begin{equation}
\frac{dV_x}{dt}= \frac{q}{m} E_x -\frac{\mu}{m} \nabla_\parallel B.
\end{equation}
We have postulated, and seen in the numerical simulations that $E_x=0$ (only $E_X$ is finite), then
\begin{equation}
\frac{dV_x}{dt}= -\frac{\mu}{m}\frac{d B}{dx} = -\frac{\mu}{m}\frac{d B}{dX},
\end{equation}
the second equality being correct when the wave magnetic perturbation is small compared to $B_0$.
This acceleration scales as
\begin{equation}
\frac{dV_x}{dt} \sim \frac{T}{m} \frac{1}{B}\frac{d B}{dX},
\end{equation}
and because the temperature is the same (in the simulations) for the ions and for the electrons, and because the electrons are lighter, this acceleration is dominant for the electrons. 

We can see that when the two waves have opposite circular polarisation, $B$ is purely time dependent. Therefore, there is no gradient, and no parallel particle acceleration at all. This case was tested with the simulation AWC011 in \citep{M2012}. The electric field in AWC011 was purely time dependent,  in accordance with Eq. (\ref{EX_general}). The electron and ion distributions $f_e(x,V_x)$ and $f_i(x,V_x)$ (not shown on a figure) remained exactly identical to their initial values, in accordance with the fact established here that there is no force acting on the parallel velocity.  

From Eq. (\ref{module_B_same_circular}), two opposite sinusoidal waves with 
the same circular polarisation make regions of acceleration with a fixed position and no time dependence. This is actually what is seen in the simulations. 
{Figs. \ref{fig_AWC008_FeXVPAR} and \ref{fig_AWC008_FpXVPAR}, 
 show the electron and ion distribution functions as functions of $X$ and $V_x$ from the simulation AWC008. We can see a periodic modulation of the electron and ion densities  with a mode number m=4 that is twice those of the two sinusoidal Aflv\'en waves. 
Concerning wave packets interactions, an example is given in Fig. \ref{fig_AWC009_CNE} 
 that show the stack plot of the electron density  for the simulation ACW009. We can notice that an important effect of the gradient of $B$ is the creation of a zone of electron depletion. This depletion  is created at the two wave packets 
crossing, but, contrarily to the $E_X$ electric field, it remains after the wave packets crossings.  The electron depletion can be seen as well on Figs.  \ref{fig_AWC009_FeXVX}, \ref{fig_AWC009_FeXVPAR} 
 that represent the electron distribution as functions of $X$ and $V_X$ and of $X$ and $V_x$. The fact of using $V_x$ or $V_X$ does not change the general aspect of the figures. This means that what is seen of these distribution is not a mere effect of projection (on the local 
or the average magnetic fields) but a real structure.  
Fig. \ref{fig_AWC009_FpXVPAR} 
 shows the ion distribution as a function of $X$ and $V_x$. We can see ion beams initiated around the positions of the wave packets crossings. Figure \ref{fig_AWC009_FpXVPAR} is more dramatic than Fig. \ref{fig_AWC008_FpXVPAR} simply because the wave packets in AWC009 have a larger amplitude than the sinusoidal waves in AWC008. }

Does the electric field play a role in the acceleration process in the perpendicular velocity $\vec V_\perp$ ? Would a projection of this velocity along the $X$ axis imply acceleration in that direction ?
Because the geometry of the problem is 1D, there is no perpendicular gradient of $B$. Because the magnetic perturbation caused by the waves is small compared to $B_0$ the magnetic field line curvature is negligible. The perpendicular velocity $\vec V_\perp$ is
therefore reduced to the sum of the cross field drift and of the polarisation drift, and its projection along the ambient magnetic field direction $\vec e_X$ is 
\begin{equation}
\vec V_\perp \cdot \vec e_X = \frac{(\vec e_X \times \vec E) \cdot \vec B}{B^2} + \frac{m}{qB^2}\frac{d( \vec E \cdot \vec e_X)}{dt} .
\end{equation}
For two sinusoidal waves with the same circular polarisation,
\begin{equation}
\vec V_\perp \cdot \vec e_X =  \frac{\omega}{k B_0^2} [B_1^2 - B_2^2  
 -4 B_1 B_2 \frac{m k V_X}{q B_0} \sin (2kX+\phi_-+n\frac{\pi}{2})]. 
 \label{eq_accel_polarisation_drift}
\end{equation} \nonumber
The two first terms come from the cross fields drift, the third from the polarisation drift.
The space dependent part of this velocity scales as
\begin{equation}
\vec V_\perp \cdot \vec e_X \sim -4 V_X \frac{B_1 B_2}{B_0^2}\frac{\omega}{\omega_c} \cos(2kX+\phi_-);
\end{equation}
it is dominant for the ions. 
In conclusion, with two circular waves with the same polarisation, a sinusoidal 
and purely space dependent ion velocity perturbation is set. It is caused by the projection $E_X$ of the electric field on the average magnetic field direction. The electrons are accelerated by a magnetic gradient force that also depends on space, and not on time. 
We can compare the ion and the electron  accelerations.
The ion acceleration is the time derivative of the expression in Eq. (\ref{eq_accel_polarisation_drift}).
We find that the ratio of the electron to ion accelerations scales as
\begin{equation}
\frac{a_e}{a_i} \sim \frac{1}{4} \frac{T_e}{T_i}\frac{m_i}{m_e} \frac{\omega_{ci}}{\omega}
\end{equation}
where we have set $V_X \sim v_{ti}$, $\mu/m \sim v_{te}^2/B_0$, and $\omega_{ci}$ is the ion cyclotron frequency. When the electron and ion temperatures are similar and $\omega/\omega_{ci} >0.1$, the electron acceleration is much higher than the ion acceleration. 

Therefore, the electrons are first accelerated. The ions do not follow and a charge separation parallel electric field is set. This electric field may accelerate the ions more efficiently than the polarisation drift would do, and cause the ion beams than we can see on Fig. \ref{fig_AWC009_FpXVPAR} and (less) on Fig.  \ref{fig_AWC008_FpXVPAR}.

The last column of Table \ref{table_simulations} indicate briefly the behaviour of the particles distribution function for each numerical simulation.
We can see that the distribution functions become easily modulated. These modulation result in density modulations and plasma cavities of moderate depth. Deep cavities, electron vortices and ion beam appear for higher wave amplitudes.

\section{Observational characteristics of APAWI} \label{sec_signature}

The APAWI process requires two counter-propagating waves of the same wavelength. Actually, it is not precisely needed that they are Alfv\'en waves, but it is necessary that, like MHD and ideal Hall MHD waves, the local electric field is perpendicular to the local magnetic field. 

From an observational point of view, we can make the distinction between the interaction of two wave packets and a standing wave. 

When two wave packets interfere, the area (along the magnetic field lines)  where $E_X$ and $\nabla B$ fields  are finite is comparable to the size of the wave packets. The duration of its occurrence is the duration of the superposition of the two wave packets. Let us assume that the wave packets have a propagation velocity $V_{wp}$ and a typical size $L_{wp}$, then, the non-oscillatory electric field last for a time lapse $\tau \sim  L_{wp}/V_{wp}$. This is the same time during which an isolated (of both) wave packet(s) is seen from a fixed position. 

For instance, in the Earth Auroral zone, a first Alfv\'en wave packet coming from the magnetotail can be reflected on the ionosphere. A second wave packet of comparable size can arrive later and cross the reflected part of the first packet. 
\citet{Louarn_1994} observed Alfv\'en wave packets in the low altitude auroral zone ($\sim 1000$ km above ground), during about 10 ms. If two wave packets like those cross each other, then, the non-oscillating electric field $E_X$ and the non-oscillating magnetic gradient  would also have a duration of approximately 10 ms. This would correspond to the duration of the acceleration process.

A standing  wave can be considered as the superposition of two sinusoidal counter-propagating waves. In that case, the region where non oscillatory $E_X$ and $\nabla B$ fields are finite extends over a significant fraction of the trapped wavelength (about 25\%). The simulations show that these regions are associated with a deep plasma density depletion.
The computation of the magnetic field modulus $B$ in Eq. (\ref{module_B_same_circular}) when the waves have the same polarisation (optimal condition for acceleration) show that it has a purely spatial dependence, with a scale-length twice smaller than those of the standing waves.  
The APAWI regions associated with trapped waves exist as long as the wave trapping occurs. This can be a much longer duration than with the interaction of two wave packets. Because of this longer duration, it may be more probable to observe APAWI associated with wave trapping than to wave packets interaction.

\section{APAWI in space plasmas} \label{sec_examples_space}


APAWI might be efficient when acceleration occurs along a closed magnetic field loop provided that parallel Alfv\'en waves propagate along it.  \cite{Melrose_2013} have discussed the role of a dense network of striated Alfv\'en waves in closed loops in the solar corona, and their link with particle acceleration. 
They suggest that  the acceleration comes from Alfv\'en waves inertial effects (implying oblique propagation). 
No explicit explanation for the oblique (quasi-perpendicular) propagation of the Alfv\'en waves is given (for the phase velocity), therefore, we may as well imagine  phase velocities parallel to the ambient magnetic field. We suggest that in the context described by \cite{Melrose_2013}, Alfv\'en waves 
with $k=k_\parallel$ 
could play the same role as the oblique inertial Alfv\'en waves relative to transport and acceleration.  
In the model in \cite{Melrose_2013}, the Alfv\'en waves transport energy from the point where they are generated toward the point where they accelerate particles and dissipate. If we suppose that coronal loops support wave packets going down from the generator region and others going up after reflection above the photosphere,  APAWI would occur at the place where they meet.

A recent example given by 
\citet{Yuan_2013}, illustrate the possibility of APAWI in the Earth radiation belts. The authors identified eight events with double-belt structure associated with magnetic storms, based on the SAMPEX data sets.  All double-belt structure events in the outer radiation belt are found during the recovery phase of a magnetic storm.  The authors suggest that the plasmapause and the strong wave-particle interactions with VLF and ULF waves near it play an important role in the development of the double-belt structures. Actually, the radiation belts are, like the loops of the sun corona, regions of closed magnetic field lines where trapped ULF can take place. These trapped waves could be intensified during the preconditions identified by \citet{Yuan_2013} and cause APAWI. Then, APAWI could be a cause of the energetic particles double-belt structure. However, it is not proved yet that APAWI can operate in this range of energies; this question is left for a further study.

Contrary to coronal loops and radiation belts, the Earth auroral zone contains open magnetic field lines. Nevertheless, 
the role played by Alfv\'en waves in auroral acceleration is of paramount importance. Alfv\'en waves are very often observed in the auroral zone \citep{Stasiewicz_2000}. 
They take the form of isolated wave packets carrying  a significant amount of Poynting flux \citep{Volwerk_1996}, as well as Alfv\'en waves trapped in an auroral zone cavity called the ionospheric Alfv\'en resonator  \citep{Chaston_2002, Lysak_2003b}. 
Both trapped waves and wave packets are associated with fluxes of accelerated particles. 
\cite{Lund_2010} have studied the cut-off of turbulent Alfv\'en wave spectra in the Earth auroral zone with FAST data. 
The authors have found cuts-off at the electron cyclotron scale, and at the electron inertial lengths. These cuts-off are the signatures of dissipation by perpendicular ion heating and by parallel electron acceleration. 
This conforms to the picture of acceleration by quasi-perpendicular Alfv\'en waves associated with electron inertial and kinetic effects \citep{Lysak_1991,Thompson_1996,Watt_2010,mottez_2011c}. 
But \cite{Lund_2010} also noticed that there is no evidence of damping at the ion inertial length. This result implies that there is no power in the quasi-parallel shear Alfv\'en mode at FAST altitudes above a few Hz. 
These waves are known to be present at higher altitudes. According to \cite{Lund_2010}, this suggest some attenuation mechanism at a plasma frame frequency of $\sim$ 1 Hz, that is, just above the cavity modes frequencies of the ionospheric Alfv\'en resonator \citep{Lysak_1991}. We suggest that the absorption of the quasi-parallel Alfv\'en waves above $\sim 1$ Hz results from APAWI associated with the trapped Alfv\'en waves in the ionospheric resonator. 

APAWI can also be an explanation to the formation of deep plasma cavities in the Earth auroral zone. These cavities where plasma density
abruptly drops to less than 50\% of the ambient plasma density are ubiquitous in the auroral zone. They are associated with auroral acceleration processes and auroral kilometric radiation \citep{Persoon_1988,Hilgers_1992b,Makela_1998}. Theories concerning their origin involve inertial Alfv\'en waves and/or ponderomotive forces \citep{Singh_1992,Shukla_1999} or transverse heating mechanism  \cite{Singh_1994}. We can see that APAWI is an alternative model for cavity formation that arises naturally when one considers the existence of trapped Alfv\'en waves in the ionospheric resonator. 

\section{Discussion and conclusion}

The APAWI process is characterised by the following properties:
The crossing of up-going and down-going parallel propagating Alfv\'en waves with the same circular polarisation generates time independent electric fields and magnetic modulus gradients along the direction of propagation of the waves. Both electric field and magnetic gradient amplitudes are quadratic non-linear effects; there is no critical threshold. They cause significant { modulation of the plasma density}. They cause { particle acceleration}.

Can APAWI be investigated with MHD ? 
The parallel electric field could be evidenced in MHD with the correct polarisation, although the electric field is rarely investigated with MHD codes. It is not specific to the PIC simulations nor to any kind of Maxwell-Vlasov analysis. But the consequences of this electric field on the plasma cannot be put clearly in evidence with MHD equations, because as shown on Fig. \ref{fig_AWC009_FpXVPAR}, 
 it does not act in a simple way on the mean velocity or the plasma temperature. The use of a plasma kinetic approach is then required, unless a proxy is found to characterise the effect of the parallel non-oscillating electric field with MHD or any other kind of fluid plasma equations. 

Many simulations have been devoted to the dynamics of incoming Aflv\'en waves and their reflected counterpart on the ionosphere, especially in the context of the ionospheric Alfv\'enic resonator \citep{Lysak_1991,Thompson_1996,Watt_2010}. 
In these situations, there are two Alfv\'en waves (whatever in the form of wave packets or of standing waves) with the same polarisation and opposite directions. Why was APAWI not seen in these simulations ?
Very often, simulations for Alfv\'en waves derive the magnetic perturbation from a vector potential $A_\parallel(X,Z,t)$ (parallel to $B_0$). This implies that $\delta \vec B$
is perpendicular to $\vec B_0$ that is compatible with the input of the theory presented in this paper. But it also means that the electric field derived from the vector potential is perpendicular to $\vec B_0$ and it does not fit the requirement that $\vec{E}$ is perpendicular to $\vec{B}=\vec B_0+ \delta \vec B$.
One of the basic principles of APAWI is forbidden by the use of $A_\parallel(X,Z,t)$. It is also important that the Alfv\'en wave polarisation is compatible with the presence of circularly polarised waves.

Of course, APAWI is not the only one acceleration process at work in space plasmas. For instance, the above cited work in  \cite{Lund_2010} also makes a large room for other Alfv\'enic acceleration processes. In order to avoid confusions, we review shortly what APAWI is not.

Ponderomotive force associated with Alfv\'en waves have been investigated by \citet{Singh_1994,Sharma_2009}. It was shown that it leads to the creation of density cavities. But ponderomotive force derives from the spatial modulation of a wave envelope. It acts therefore at the same scale as the envelope and not at the smaller scale defined by the wavelength. In APAWI, as we have seen with the theoretical treatment of two waves interaction, the acceleration structures scale as the wavelength, and they do not involve any modulation of the waves amplitudes. 
Therefore APAWI is not acceleration by a ponderomotive force. This is an alternative process acting on a smaller scale length.

Because the cavities created by the APAWI are smaller than the wavelengths, the waves that caused these cavities cannot be trapped in it. But it is not excluded that waves with smaller wavelengths then become trapped in these cavities formed by APAWI. Then, the observational data may become more complex, making its analysis more difficult. 

Contrary to acceleration by inertial of kinetic Alfv\'en waves, no oblique propagation is required, and instead of $k_\perp c /\omega_{pe} \sim 1$ or $k_\perp \rho_i \sim 1$, it is simply required that $k_\perp=0$. There is no specific requirement on the parallel wavelength. 

The acceleration by an Alfv\'en wave with a velocity close to the thermal electron velocity has been investigated by \citet{Watt_2008}. This involves Landau resonance for electrons in the wave fields.
With APAWI, the acceleration region is static or quasi-static. No resonance effect that would select the range of initial energies of the accelerated particles is involved. This is not resonant acceleration. 

 An  Alfv\'en wave in parallel propagation over plasma density cavities soon develop small transverse scales, parallel electric fields \citep{genot_1999} and causes plasma acceleration \citep{Mottez_2001_b}. With APAWI, the density cavities are a result of the process and not an initial ingredient. 
APAWI can work in an initially uniform plasma. 

{The decay of Alfv\'en waves into a daughter Alfv\'en wave and an acoustic-like wave can trigger the formation of ion beams \citep{Matteini_2010}. Modulational  instabilities can result as well in ion beam acceleration \citep{Araneda_2008}. Is ion acceleration
seen with APAWI really the result of the process described in section \ref{sec_acceleration}, or the result of an unnoticed AW decay or modulational instability ? 
Figure \ref{fig_AWC008_FpXVPAR} shows that the modulation of the ion distribution function has a wave number $m=4$ for initial Alfv\'en wave  numbers $m_1=m_2=2$. This is consistent with the multiplication by two of the initial wave vector seen, for instance, in Eq. (\ref{module_B_same_circular}). This is not compatible with the slightly more complex wave number ratios involved with parametric instabilities. Therefore, we are tempted to conclude that ion acceleration is directly related to APAWI, and not to a parametric instability.}

Therefore we can conclude that APAWI does not involve ponderomotive force, inertial or kinetic Alfv\'en waves, resonant acceleration or  propagation effects on a non-uniform plasma, or parametric instabilities. Of course, APAWI can act simultaneously with one of these and the consequences of APAWI can be similar to those of other acceleration mechanisms. For instance, like ponderomotive forces, APAWI causes deep plasma density depletions; {like the parametric instabilities, it can cause ion beams}.


In conclusion of this paper, acceleration by parallel Alfv\'en waves interaction (APAWI) is a new method of accelerating plasma particles with Alfv\'en waves. The existence of APAWI does not rule out the relevance of other Alfv\'enic processes, but it enlarges the range of plasma parameters and waves polarisations that make possible plasma acceleration by Alfv\'en waves. More specifically, it is the only acceleration process known up to now that involves only parallel Alfv\'en waves and an initially homogeneous plasma.  APAWI may be relevant in a very large number of space plasma contexts, including planetary magnetospheres and magnetic loops in the Sun corona. It is a very efficient way of creating plasma cavities such as those observed in the Earth auroral zone.

%


\appendix
\section{Parallel electric field}   \label{sec_faraday}

We define the two angles $\theta_Y$ and $\theta_Z$ by the relations
\begin{equation}
\frac{B_Y}{B_0}=\tan \theta_Y   \mbox{ and } \frac{B_Z}{B_0}=\tan \theta_Z.
\end{equation}
Because we consider that the MHD waves amplitude is low relative to the ambient magnetic field $B_0$, we consider that these angles are small.

\begin{figure}
\vspace*{2mm}
\begin{center}
\includegraphics[width=8.8cm]{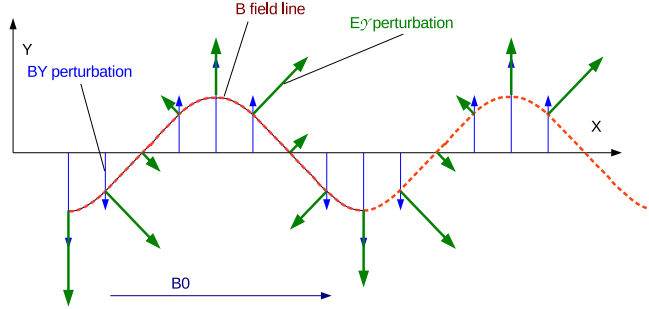} 
\end{center}
\caption{Schematic explanation of the parallel electric field $E_X$ parallel to the 
average magnetic field $\vec{B_0}$.}
\label{fig_N_L_polar_BE}
\end{figure}

The ideal Ohm's law, used for MHD waves, implies that the electric field and the magnetic field are orthogonal. This holds when the Ohm's law includes the Hall effect, 
 \begin{equation}
 \vec E + \vec v \times \vec B = \frac{\vec J}{qn} \times \vec B.
 \end{equation}
 The situation is schematically displayed  on Fig. \ref{fig_N_L_polar_BE} 
 where the
 electric field  is represented by thick arrows (green in the electronic version). The electric field is perpendicular to the (dashed/red) magnetic field line, and the magnetic perturbation vectors (thin blue arrows) are perpendicular to the $X$ direction. In terms of vector components, $E_x=0$, but we can see that $E_X$ (projection of the electric field along the \textit{average} magnetic field $\vec{B_0}$) can be finite.
 The local coordinate system $(x,y,z)$ is based on the direction of the magnetic field, with $\vec{e_x}$ parallel to $\vec{B_0}$. 
%
%
%
The magnetic field lines are defined by,
\begin{equation}
\frac{dX}{B_0}=\frac{dY}{B_Y}=\frac{dZ}{B_Z}.
\end{equation}

The following computation is based on the hypothesis of small angles $\theta_Y \ll 1$ and $\theta_Z \ll 1$. It is based on first order developments relative to these angles. 
Then,
\begin{equation} \label{theta_Y_de_B}
\frac{dY}{dX}=\frac{B_Y}{B_0}=\tan \theta_Y  = \theta_Y \; \mbox{ and } \; \frac{dZ}{dX}=\frac{B_Z}{B_0}=\tan \theta_Z = \theta_Z.
\end{equation}
Making projections, 
\begin{equation} \label{eq_projection_EX}
E_X = -E_y \sin \theta_Y -E_z \sin \theta_Z \sim -E_y \frac{B_Y}{B_0}  -E_z \frac{B_Z}{B_0}.
\end{equation}

With two sinusoidal waves labelled $1$ and $2$, 
\begin{eqnarray} 
E_x &=& 0, \label{eq_E}\\
\nonumber
E_y &=& E_{1y} \cos(\omega t - kX +\phi_{E_{1y}})+
E_{2y} \cos(\omega t + kX +\phi_{E_{2y}}),\\   
\nonumber
E_z &=& E_{1z} \cos(\omega t - kX +\phi_{E_{1z}})+
E_{2z} \cos(\omega t + kX +\phi_{E_{2z}}).   
\end{eqnarray}
In addition to the computation presented in \citep{M2012}, we take into account 
the constraint raised by the  Faraday equation
\begin{equation}
\nabla \times \vec E = -\frac{\partial \vec B}{\partial t}.
\end{equation}

At first order in $\theta_Y$ and $\theta_Z$, 
\begin{eqnarray} \label{e_y_de_e_Y}
\vec{e_y}= \vec{e_Y} - \theta_Y \vec{e_X}  \mbox{, } \; 
\vec{e_z}= \vec{e_Z} - \theta_Z \vec{e_X}  \\ \nonumber
\vec{e_x}= \theta_Y \vec{e_Y} + \theta_Z \vec{e_Z} + \vec{e_X}.
\end{eqnarray}
Reciprocally,
\begin{eqnarray}
\vec{e_Y}= \vec{e_y} + \theta_Y \vec{e_x}  \mbox{, } \; 
\vec{e_Z}= \vec{e_z} + \theta_Z \vec{e_x}  \\ \nonumber
\vec{e_X}= -\theta_Y \vec{e_y} - \theta_Z \vec{e_z} + \vec{e_x}.
\end{eqnarray}
The above expression of $\vec{e_X}$ proves the validity of Eq. (\ref{eq_projection_EX}).
Because the electromagnetic field depends on $X$, the Faraday law is written in the 
$(\vec{e_X},\vec{e_Y},\vec{e_Z})$ basis,
\begin{eqnarray}
\frac{\partial B_Y}{\partial t}= \frac{\partial E_Z}{\partial X}  \mbox{ and } \; 
\frac{\partial B_Z}{\partial t}= -\frac{\partial E_Y}{\partial X}  \mbox{ and } \; 
\frac{\partial B_X}{\partial t} = 0. 
\end{eqnarray}
To the first order in $\theta_Y$ and $\theta_Z$, 
\begin{equation}
E_Y=\vec{E} \cdot \vec{e_Y}= E_y \; \mbox{ and } \;
E_Z=\vec{E} \cdot \vec{e_Z}= E_z.
\end{equation}
Therefore,
\begin{equation}
\frac{\partial B_Y}{\partial t}= \frac{\partial E_z}{\partial X} \; \mbox{ and } \;
\frac{\partial B_Z}{\partial t}= -\frac{\partial E_y}{\partial X} 
\end{equation}
Considering Eqs. (\ref{eq_B},\ref{eq_E}), the solution is 
\begin{eqnarray}
\omega B_{1Y}= - k E_{1z}   \, \mbox{ and } \, \phi_{B_{1Y}}= \phi_{E_{1z}} \mbox{ and } \; 
\omega B_{1Z}= + k E_{1y}   \, \mbox{ and } \, \phi_{B_{1Z}}= \phi_{E_{1y}}, \\ \nonumber
\omega B_{2Y}= - k E_{2z}   \, \mbox{ and } \, \phi_{B_{2Y}}= \phi_{E_{2z}} \mbox{ and } \; 
\omega B_{2Z}= + k E_{2y}   \, \mbox{ and } \, \phi_{B_{2Z}}= \phi_{E_{2y}},
\end{eqnarray} 
and this is equivalent to
\begin{eqnarray}
\nonumber
E_{1y}= +\frac{\omega B_{1Z}}{k}   \, \mbox{ and } \, \phi_{E_{1y}}=\phi_{B_{1Z}} \mbox{ and } \; 
E_{1z}= -\frac{\omega B_{1Y}}{k}   \, \mbox{ and } \, \phi_{E_{1z}}=\phi_{B_{1Y}}, \\ \nonumber
E_{2y}= -\frac{\omega B_{2Z}}{k}   \, \mbox{ and } \, \phi_{E_{2y}}=\phi_{B_{2Z}} \mbox{ and } \; 
E_{2z}= +\frac{\omega B_{2Y}}{k}   \, \mbox{ and } \, \phi_{E_{2z}}=\phi_{B_{2Y}}.  \label{eq_de_Faraday}
\end{eqnarray} 
Then,
\begin{eqnarray} 
\nonumber
E_y &=& +\frac{\omega B_{1Z}}{k} \cos(\omega t - kX +\phi_{B_{1Z}})-
\frac{\omega B_{2Z}}{k} \cos(\omega t + kX +\phi_{B_{2Z}}),\\   
E_z &=& -\frac{\omega B_{1Y}}{k} \cos(\omega t - kX +\phi_{B_{1Y}})+
\frac{\omega B_{2Y}}{k} \cos(\omega t + kX +\phi_{B_{2Y}}). \label{eq_E_bis}
\end{eqnarray}


With Eqs. (\ref{eq_B},\ref{eq_E},\ref{eq_de_Faraday}), the projection in Eq.  (\ref{eq_projection_EX}) provides Eqs. (\ref{EX_general}).

\begin{acknowledgements}
The numerical simulations were performed at the computing center (DIO) of the Paris-Meudon observatory. Financial help was supplied for this project by CNRS/INSU/PNST (Programme National Soleil Terre). 
\end{acknowledgements}

\bibliographystyle{jpp} 
\bibliography{article} 

\begin{thebibliography}{33}
\expandafter\ifx\csname natexlab\endcsname\relax\def\natexlab#1{#1}\fi

\bibitem[{Araneda} {\em et~al.\/}(2008){Araneda}, {Marsch} \&
  {F.-Vi{\~n}as}]{Araneda_2008}
{\sc {Araneda}, J.~A., {Marsch}, E. \& {F.-Vi{\~n}as}, A.} 2008 {Proton Core
  Heating and Beam Formation via Parametrically Unstable Alfv{\'e}n-Cyclotron
  Waves}. {\em Physical Review Letters\/} {\bf 100}~(12), 125003.

\bibitem[{Buti} {\em et~al.\/}(2000){Buti}, {Velli}, {Liewer}, {Goldstein} \&
  {Hada}]{Buti_2000}
{\sc {Buti}, B., {Velli}, M., {Liewer}, P.~C., {Goldstein}, B.~E. \& {Hada},
  T.} 2000 {Hybrid simulations of collapse of Alfv{\'e}nic wave packets}. {\em
  Physics of Plasmas\/} {\bf 7}, 3998--4003.

\bibitem[{Chaston} {\em et~al.\/}(2002){Chaston}, {Bonnell}, {Carlson},
  {Berthomier}, {Peticolas}, {Roth}, {McFadden}, {Ergun} \&
  {Strangeway}]{Chaston_2002}
{\sc {Chaston}, C.~C., {Bonnell}, J.~W., {Carlson}, C.~W., {Berthomier}, M.,
  {Peticolas}, L.~M., {Roth}, I., {McFadden}, J.~P., {Ergun}, R.~E. \&
  {Strangeway}, R.~J.} 2002 {Electron acceleration in the ionospheric Alfven
  resonator}. {\em Journal of Geophysical Research (Space Physics)\/} {\bf
  107}~(11), 41--1.

\bibitem[{G{\'e}not} {\em et~al.\/}(1999){G{\'e}not}, {Louarn} \& {Le
  Qu{\'e}au}]{genot_1999}
{\sc {G{\'e}not}, V., {Louarn}, P. \& {Le Qu{\'e}au}, D.} 1999 {A study of the
  propagation of Alfv{\'e}n waves in the auroral density cavities}. {\em
  Journal of Geophysical Research (Space Physics)\/} {\bf 104}~(13),
  22649--22656.

\bibitem[{G{\'e}not} {\em et~al.\/}(2001){G{\'e}not}, {Mottez} \&
  {Louarn}]{Mottez_2001_b}
{\sc {G{\'e}not}, V., {Mottez}, F. \& {Louarn}, P.} 2001 {Particle Acceleration
  Linked to Alfven Wave Propagation on Small Scale Density Gradients}. {\em
  Physics and Chemistry of the Earth C\/} {\bf 26}, 219--222.

\bibitem[{Goertz}(1984)]{Goertz_1984}
{\sc {Goertz}, C.~K.} 1984 {Kinetic Alfv{\' e}n waves on auroral field lines}.
  {\em Planetary and Space Science\/} {\bf 32}, 1387--1392.

\bibitem[{Hasegawa} \& {Mima}(1978)]{Hasegawa_1978}
{\sc {Hasegawa}, A. \& {Mima}, K.} 1978 {Anomalous transport produced by
  kinetic Alfven wave turbulence}. {\em Journal of Geophysical Research (Space
  Physics)\/} {\bf 83}~(12), 1117--1123.

\bibitem[{Hilgers}(1992)]{Hilgers_1992b}
{\sc {Hilgers}, A.} 1992 {The auroral radiating plasma cavities}. {\em
  Geophysical Research Letters\/} {\bf 19}, 237--240.

\bibitem[{Knudsen}(2001)]{Knudsen_2001}
{\sc {Knudsen}, D.~J.} 2001 {Structure, Acceleration, and Energy in Auroral
  Arcs and the Role of Alfv{\'e}n Waves}. {\em Space Science Review\/} {\bf
  95}, 501--511.

\bibitem[{Louarn} {\em et~al.\/}(1994){Louarn}, {Wahlund}, {Chust}, {de
  Feraudy}, {Roux}, {Holback}, {Dovner}, {Eriksson} \& {Holmgren}]{Louarn_1994}
{\sc {Louarn}, P., {Wahlund}, J.~E., {Chust}, T., {de Feraudy}, H., {Roux}, A.,
  {Holback}, B., {Dovner}, P.~O., {Eriksson}, A.~I. \& {Holmgren}, G.} 1994
  {Observation of kinetic Alfven waves by the Freja spacecraft}. {\em Geophys.
  Res. Lett.\/} {\bf 21}, 1847--+.

\bibitem[{Lund}(2010)]{Lund_2010}
{\sc {Lund}, E.~J.} 2010 {On the dissipation scale of broadband ELF waves in
  the auroral region}. {\em Journal of Geophysical Research (Space Physics)\/}
  {\bf 115}, 1201.

\bibitem[{Lysak}(1991)]{Lysak_1991}
{\sc {Lysak}, R.~L.} 1991 {Feedback instability of the ionospheric resonant
  cavity}. {\em Journal of Geophysical Research (Space Physics)\/} {\bf 96},
  1553--1568.

\bibitem[{Lysak} \& {Song}(2003{\natexlab{{\em a\/}}})]{Lysak_2003b}
{\sc {Lysak}, R.~L. \& {Song}, Y.} 2003{\natexlab{{\em a\/}}} {Kinetic theory
  of the Alfv{\' e}n wave acceleration of auroral electrons}. {\em Journal of
  Geophysical Research (Space Physics)\/} {\bf 108}~(4), 6--1.

\bibitem[{Lysak} \& {Song}(2003{\natexlab{{\em b\/}}})]{Lysak_2003}
{\sc {Lysak}, R.~L. \& {Song}, Y.} 2003{\natexlab{{\em b\/}}} {Nonlocal kinetic
  theory of Alfv{\'e}n waves on dipolar field lines}. {\em Journal of
  Geophysical Research (Space Physics)\/} {\bf 108}, 1327--+.

\bibitem[{M{\"a}kel{\"a}} {\em et~al.\/}(1998){M{\"a}kel{\"a}}, {M{\"a}lkki},
  {Koskinen}, {Boehm}, {Holback} \& {Eliasson}]{Makela_1998}
{\sc {M{\"a}kel{\"a}}, J.~S., {M{\"a}lkki}, A., {Koskinen}, H., {Boehm}, M.,
  {Holback}, B. \& {Eliasson}, L.} 1998 {Observations of mesoscale auroral
  plasma cavity crossings with the Freja satellite}. {\em Journal of
  Geophysical Research (Space Physics)\/} {\bf 103}, 9391--9404.

\bibitem[{Matteini} {\em et~al.\/}(2010){Matteini}, {Landi}, {Velli} \&
  {Hellinger}]{Matteini_2010}
{\sc {Matteini}, L., {Landi}, S., {Velli}, M. \& {Hellinger}, P.} 2010
  {Kinetics of parametric instabilities of Alfv{\'e}n waves: Evolution of ion
  distribution functions}. {\em Journal of Geophysical Research (Space
  Physics)\/} {\bf 115}, 9106.

\bibitem[{Melrose} \& {Wheatland}(2013)]{Melrose_2013}
{\sc {Melrose}, D.~B. \& {Wheatland}, M.~S.} 2013 {Transfer of Energy,
  Potential, and Current by Alfv{\'e}n Waves in Solar Flares}. {\em Solar
  Physics\/} {\bf 288}, 223--240.

\bibitem[{Mottez}(2008)]{mottez_2008_a}
{\sc {Mottez}, F.} 2008 {A guiding centre direct implicit scheme for magnetized
  plasma simulations}. {\em Journal of Computational Physics\/} {\bf 227},
  3260--3281.

\bibitem[{Mottez}(2012{\natexlab{{\em a\/}}})]{M2012}
{\sc {Mottez}, F.} 2012{\natexlab{{\em a\/}}} {Non-propagating electric and
  density structures formed through non-linear interaction of Alfv{\'e}n
  waves}. {\em Annales Geophysicae\/} {\bf 30}, 81--95.

\bibitem[{Mottez}(2012{\natexlab{{\em b\/}}})]{mottez_2012b}
{\sc {Mottez}, F.} 2012{\natexlab{{\em b\/}}} {The role Alfv\'en waves in the
  generation of Earth polar auroras}. {\em Proceedings of "Waves and
  Instabilities in Space and Astrophysical Plasmas" (WISAP) Eilat, Israel, June
  19th - June 24th, 2011\/} .

\bibitem[{Mottez} {\em et~al.\/}(1998){Mottez}, {Adam} \& {Heron}]{Mottez_1998}
{\sc {Mottez}, F., {Adam}, J.~C. \& {Heron}, A.} 1998 {A new guiding centre PIC
  scheme for electromagnetic highly magnetized plasma simulation}. {\em
  Computer Physics Communications\/} {\bf 113}, 109--130.

\bibitem[{Mottez} \& {G\'enot}(2011)]{mottez_2011c}
{\sc {Mottez}, F. \& {G\'enot}, V.} 2011 {Electron acceleration by an
  Alfv\'enic pulse propagating in an auroral plasma cavity}. {\em Journal of
  Geophysical Research\/} {\bf 116}, A00K15.

\bibitem[{Persoon} {\em et~al.\/}(1988){Persoon}, {Gurnett}, {Peterson},
  {Waite}, {Burch} \& {Green}]{Persoon_1988}
{\sc {Persoon}, A.~M., {Gurnett}, D.~A., {Peterson}, W.~K., {Waite}, Jr.,
  J.~H., {Burch}, J.~L. \& {Green}, J.~L.} 1988 {Electron density depletions in
  the nightside auroral zone}. {\em Journal of Geophysical Research (Space
  Physics)\/} {\bf 93}, 1871--1895.

\bibitem[{Sharma} \& {Singh}(2009)]{Sharma_2009}
{\sc {Sharma}, R.~P. \& {Singh}, H.~D.} 2009 {Density cavities associated with
  inertial Alfv{\'e}n waves in the auroral plasma}. {\em Journal of Geophysical
  Research (Space Physics)\/} {\bf 114}, 3109.

\bibitem[{Shukla} \& {Stenflo}(1999)]{Shukla_1999}
{\sc {Shukla}, P.~K. \& {Stenflo}, L.} 1999 {Plasma density cavitation due to
  inertial Alfv{\'e}n wave heating}. {\em Physics of Plasmas\/} {\bf 6},
  4120--4122.

\bibitem[{Singh}(1992)]{Singh_1992}
{\sc {Singh}, N.} 1992 {Plasma perturbations created by transverse ion heating
  events in the magnetosphere}. {\em Journal of Geophysical Research (Space
  Physics)\/} {\bf 97}, 4235--4249.

\bibitem[{Singh}(1994)]{Singh_1994}
{\sc {Singh}, N.} 1994 {Pondermotive versus mirror force in creation of the
  filamentary cavities in auroral plasma}. {\em Geophys. Res. Lett.\/} {\bf
  21}, 257--260.

\bibitem[{Stasiewicz} {\em et~al.\/}(2000){Stasiewicz}, {Bellan}, {Chaston},
  {Kletzing}, {Lysak}, {Maggs}, {Pokhotelov}, {Seyler}, {Shukla}, {Stenflo},
  {Streltsov} \& {Wahlund}]{Stasiewicz_2000}
{\sc {Stasiewicz}, K., {Bellan}, P., {Chaston}, C., {Kletzing}, C., {Lysak},
  R., {Maggs}, J., {Pokhotelov}, O., {Seyler}, C., {Shukla}, P., {Stenflo}, L.,
  {Streltsov}, A. \& {Wahlund}, {J.-E.}} 2000 {Small Scale Alfv{\'e}nic
  Structure in the Aurora}. {\em Space Science Review\/} {\bf 92}, 423--533.

\bibitem[{Thompson} \& {Lysak}(1996)]{Thompson_1996}
{\sc {Thompson}, B.~J. \& {Lysak}, R.~L.} 1996 {Electron acceleration by
  inertial Alfv{\'e}n waves}. {\em Journal of Geophysical Research (Space
  Physics)\/} {\bf 101}, 5359--5370.

\bibitem[{Volwerk} {\em et~al.\/}(1996){Volwerk}, {Louarn}, {Chust}, {Roux},
  {de Feraudy} \& {Holback}]{Volwerk_1996}
{\sc {Volwerk}, M., {Louarn}, P., {Chust}, T., {Roux}, A., {de Feraudy}, H. \&
  {Holback}, B.} 1996 {Solitary kinetic Alfv{\' e}n waves: A study of the
  Poynting flux}. {\em Journal of Geophysical Research (Space Physics)\/} {\bf
  101}~(10), 13335--13344.

\bibitem[{Watt} \& {Rankin}(2008)]{Watt_2008}
{\sc {Watt}, C.~E.~J. \& {Rankin}, R.} 2008 {Electron acceleration and parallel
  electric fields due to kinetic Alfv{\'e}n waves in plasma with similar
  thermal and Alfv{\'e}n speeds}. {\em Advances in Space Research\/} {\bf 42},
  964--969.

\bibitem[{Watt} \& {Rankin}(2010)]{Watt_2010}
{\sc {Watt}, C.~E.~J. \& {Rankin}, R.} 2010 {Do magnetospheric shear Alfv{\'e}n
  waves generate sufficient electron energy flux to power the aurora?} {\em
  Journal of Geophysical Research (Space Physics)\/} {\bf 115}~(14), 7224--+.

\bibitem[{Yuan} \& {Zong}(2013)]{Yuan_2013}
{\sc {Yuan}, C. \& {Zong}, Q.} 2013 {The double-belt outer radiation belt
  during CME- and CIR-driven geomagnetic storms}. {\em Journal of Geophysical
  Research (Space Physics)\/} {\bf 118}, 6291--6301.

\end{thebibliography}

\end{document}